\documentclass{siamart1116}

\usepackage{amssymb}

\usepackage{tikz}

\usetikzlibrary{arrows,shapes}

\usepackage{booktabs} % Required for the top and bottom rules in the table

\usepackage{float} % Required for specifying the exact location of a figure or table
\usepackage{graphicx} % Required for including images

\usepackage{amsmath} % Required for the use of align

\usepackage{subfig}

\usepackage{fancyhdr}

\usepackage{bm}% bold math

\usepackage[linewidth=1pt]{mdframed}

\usepackage{lipsum}

\usepackage[toc,page]{appendix} % The appendix package

\title{A Time-Spectral Method for Initial-Value Problems Using a Novel Spatial Subdomain Scheme}

\author{Kristoffer Lindvall%
  \thanks{Department of Fusion Plasma Physics, KTH Royal Institute of Technology, Stockholm, Sweden. (\email{kristoffer.lindvall@ee.kth.se}, \email{jan.scheffel@ee.kth.se})}%
  \and
  Jan Scheffel%
  \footnotemark[1]
}
\headers{A Time-Spectral Method}
{Kristoffer Lindvall and Jan Scheffel}

\begin{document}
\raggedbottom
\maketitle

\setlength{\intextsep}{10pt}
%\begin{tcbverbatimwrite}{tmp_\jobname_abstract.tex}
\begin{abstract}
We analyse a new subdomain scheme for a time-spectral method for solving initial boundary value problems. Whilst spectral methods are commonplace for spatially dependent systems, finite difference schemes are typically applied for the temporal domain. The Generalized Weighted Residual Method (GWRM) is a fully spectral method in that it spectrally decomposes all specified domains, including the temporal domain, with multivariate Chebyshev polynomials. The  Common Boundary-Condition method (CBC) is a spatial subdomain scheme that solves the physical equations independently from the global connection of subdomains. It is here evaluated against two finite difference methods. For the linearised Burger equation the CBC-GWRM is $\sim30$\% faster and $\sim50$\% more memory efficient than the semi implicit Crank-Nicolson method at a maximum error $\sim10^{-5}$. For a forced wave equation the CBC-GWRM manages to average efficiently over the small time-scale in the entire temporal domain. The CBC-GWRM is also applied to the linearised ideal magnetohydrodynamic (MHD) equations for a screw pinch equilibrium. The growth rate of the most unstable mode was efficiently computed with an error $<0.1\%$.

\end{abstract}

%\end{tcbverbatimwrite}
%\input{tmp_\jobname_abstract.tex}


\section{Introduction}
In the field of fusion plasma physics complex sets of linear PDEs need frequently be solved. A particular example is the sub-discipline devoted to the analysis of linearised magnetohydrodynamic equilibria and the evolution of unstable modes \cite{Bateman:1,Furth:1,Glasser:1,Scheffel:GWRM1,Strauss:1}. The analysis consists of linearising the governing magnetohydrodynamic equations, assuming an equilibrium, introducing a perturbation, and then calculating the mode growth rate. The existence of multiple time scales require many time steps to resolve. It would be beneficial to average over small scale dynamics efficiently in order to obtain the dominant unstable mode. 

The efforts of increasing the efficiency and accuracy of time dependent numerical schemes are long standing. Authors such as Y. Morchoisne (1979) \cite{Morchoisne:1}, Peryet. et al. (1983) \cite{Peyret:1}, Tal-Ezer (1983-84) \cite{Tal-Ezer:1,Tal-Ezer:2}, and  Bar-Yoseph et. al. (1995) \cite{Bar-Yoseph:1}, to name a few, have pioneered the concept of time-spectral methods. A number of pseudo-spectral methods with polynomial basis functions such as Fourier, Legendre, and Chebyshev polynomials, with their respective advantages have been analyzed by the various authors. In the works by P. Dutt \cite{Dutt:1} and Maerschalck et. al. \cite{Maerschalck:1} a space-time Chebyshev collocation method was implemented to solve initial value problems in a least-squared sense. It is in this vein that the current work is pursued.

The time-spectral method proposed in \cite{Scheffel:GWRM2} is a fully spectral method that spectrally decomposes all domains including spatial, temporal, and parameter domains. Multivariate Chebyshev polynomials are used to represent all domains in a solution ansatz. The time-spectral method \cite{Scheffel:GWRM2} executes all computations in spectral space, i.e. no collocation points are used. The only spatial points that are taken into consideration are the points where the spatial subdomains overlap. The spatial subdomains are overlapped for the purpose of having a two-point contact. This subdomain method ensures spectral accuracy in the entire domain. Unfortunately, the resulting global matrix equations associated with solving the algebraic equations for the solution coefficients may become excessively large \cite{Boyd:1,Canuto:1,Kopriva:1}. 

There are several popular methods for dealing with this issue. One such method is to use parallel algorithms that solve for large sparse matrices \cite{Mittal:1,Spagnoli:1}, and another is to create independent spatial domains that can be solved locally with boundary-solvers that connect the subdomains \cite{Canuto:1,Giraldo:1,Kopriva:1,Peraire:1,Scheffel:GWRM3}. The idea of reducing the global degrees of freedom was first implemented with the static condensation method \cite{Guyan:1}, and thereafter has seen many variations that make the algorithm more general and dynamic \cite{Qu:1}. The static condensation algorithm has been popular with finite element methods, however similar algorithms have been introduced to spectral methods. The influence matrix technique is such a condensation algorithm related to spectral methods \cite{Boguslawski:1,Vanel:1}. In \cite{Boguslawski:1} the Navier-Stokes streamfunction-vorticity equations are solved with an iterative domain decomposition method by \cite{Louchart:1} in conjunction with the influence matrix technique. The influence matrix uses the fact that the vorticity and the stream function are linear so that they can be decomposed into a time-dependent term and a static term. The influence matrix includes the static terms, which then enforce the no-slip conditions on the boundaries in every finite time-step.

This brings us to the current spatial subdomain scheme termed the Common Boundary-Condition (CBC) Method. This method solves the "private" physical equations locally and separately, but the "boundary" equations that connect the subdomains are solved globally. This will in effect reduce the global degrees of freedom a substantial amount. The CBC method, compared to the other methods, is a general subdomain scheme for the time-spectral method by calculating how the external boundary conditions and internal patching conditions influence the private, physical equations. Thus, the CBC method does not decompose patching conditions, instead it calculates implicit and explicit derivatives of all patching conditions with respect to the private equations in the individual subdomains. The CBC method also avoids any matrix singularity issues by using a modal representation, instead of avoiding the corner points of the subdomains \cite{Boguslawski:1}. Most importantly, it takes into account the temporal modes, and adjusts the boundary and patching conditions throughout the entire temporal domain.

The paper is organized as follows. Section 2 briefly summarizes the Generalized Weighted Residual Method (GWRM), which is the foundation of the time-spectral method. The new spatial subdomain scheme that has been implemented is also laid out. Next follows solution of test problems in section 3. The test problems include the linearised 1D Burger equation, a forced wave equation and a case relating to the linearised ideal MHD equations. Section 4 contains a discussion, including some observations, possible objections, and resolutions. The paper closes with a conclusion.

\section{Method}
We will here present a brief review of the Generalized Weighted Residual Method GWRM. For a full and detailed presentation the reader is directed to \cite{Scheffel:GWRM2}.

\subsection{Weighted residual formulation}
We start by assuming a set of partial differential equations,
\begin{align}
\frac{{\partial}\textbf{u}}{{\partial}t}=D\textbf{u} + f.
\label{eq:M1}
\end{align}
where $D$ is an arbitrary linear or non-linear operator and $f(t,\textbf{x};\textbf{p})$ is a known force term. The solution is approximated with multivariate Chebyshev expansion series in time, space, and parameter space. For a single spatial dimension and one parameter we have
\begin{align}
u(t,x;p)~{\approx}~\text{U}(\tau,\xi;P)=\sum^{K}_{k=0}{}^{\prime}\sum^{L}_{l=0}{}^{\prime}\sum^{\Lambda}_{{\lambda}=0}{}^{\prime}a_{kl\lambda}T_{k}(\tau)T_{l}(\xi)T_{\lambda}(P).
\label{eq:M2}
\end{align}
Here ($'$) denotes that the first term in the summation is divided by two. Since Chebyshev polynomials are defined in the range $[-1,1]$, a change of variables is performed; $\sigma=(z-A_z)/B_z$, $A_z=(z_1+z_0)/2$, and $B_z=(z_1-z_0)/2$. Here $\sigma$ signifies the transformed variable and $z$ is the physical variable, with indices 0 and 1 denoting domain limits.

The weighted residual formulation is obtained by substituting \eqref{eq:M2} into \eqref{eq:M1}, multiplying the residual with weight functions, and integrating over the entire domain;
\begin{align}
\int^{p_1}_{p_0}\int^{x_1}_{x_0}\int^{t_1}_{t_0}RT_{q}(\tau)T_{r}(\xi)T_{s}(P)\text{w}_t\text{w}_x\text{w}_pdtdxdp=0.
\label{eq:M3}
\end{align}
The residual has the form
\begin{align}
R=u(t,x;p)-[u(t_0,x;p)+\int^{t}_{t_0}(Du+f)dt^{\prime}],
\label{eq:M4}
\end{align}
and the weight function,
\begin{align}
\text{w}_z=(1-\sigma^2)^{-1/2}.
\label{eq:M5}
\end{align}
The resulting algebraic equations may be written as
\begin{align}
a_{qrs}=2\delta_{q0}b_{rs}+A_{qrs}+F_{qrs}
\label{eq:M6}
\end{align}
This simple and general formula contains Chebyshev coefficients relating to the initial conditions $b_{rs}$, the linear/non-linear operator term $A_{qrs}$, and the force term $F_{qrs}$. Boundary equations take the place of the highermost spatial modes of $a_{qrs}$ after inserting Eq. (2) into the given external boundary conditions; internal boundary conditions are produced by coupling the spatial subdomains. In the GWRM equation \eqref{eq:M6} plus boundary condition equations are solved with a semi-implicit root solver (SIR) \cite{Scheffel:SIR}. This may entail solving a global matrix equation for all the $(K+1)(L+1)(\Lambda+1)$ coefficients. The number of numerical operations, and the CPU time, would then scale as $((K+1)(L+1)(\Lambda+1))^3$ and the memory requirements as $((K+1)(L+1)(\Lambda+1))^2$. This is of course unacceptable from a viewpoint of efficiency. 

\subsection{Subdomain scheme}
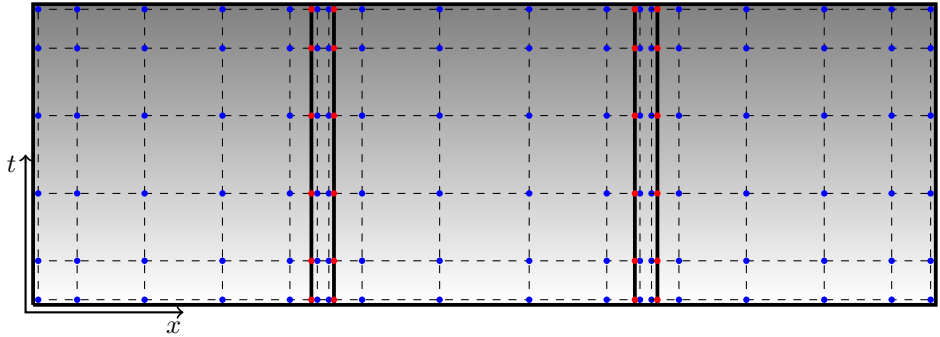
\begin{figure}[H]
\begin{center}
\begin{tikzpicture}[scale=2]

\draw [<->,thick] (-0.05,1) node (yaxis) [below] {}
        |- (1,-0.05) node (xaxis) [left] {};
\draw (yaxis) node[left] {$t$}
(xaxis) node[below] {$x$};
        
\path[shade,draw] (6,0) rectangle (0,2);
%% Vertical lines 1
\draw [dashed] (0.340741737e-1,0) -- (0.340741737e-1,2);
\draw [dashed] (.2928932190,0) -- (.2928932190,2);
\draw [dashed] (.7411809549,0) -- (.7411809549,2);
\draw [dashed] (1.258819045,0) -- (1.258819045,2);
\draw [dashed] (1.707106781,0) -- (1.707106781,2);
\draw [dashed] (1.965925826,0) -- (1.965925826,2);
%% Horizontal lines 1
\draw [dashed] (0,0.340741737e-1) -- (2,0.340741737e-1);
\draw [dashed] (0,.2928932190) -- (2,.2928932190);
\draw [dashed] (0,.7411809549) -- (2,.7411809549);
\draw [dashed] (0,1.258819045) -- (2,1.258819045);
\draw [dashed] (0,1.707106781) -- (2,1.707106781);
\draw [dashed] (0,1.965925826) -- (2,1.965925826);

\fill[blue] (0.340741737e-1,0.340741737e-1) circle (0.6pt) node {};
\fill[blue] (0.340741737e-1,.2928932190) circle (0.6pt) node {};
\fill[blue] (0.340741737e-1,.7411809549) circle (0.6pt) node {};
\fill[blue] (0.340741737e-1,1.258819045) circle (0.6pt) node {};
\fill[blue] (0.340741737e-1,1.707106781) circle (0.6pt) node {};
\fill[blue] (0.340741737e-1,1.965925826) circle (0.6pt) node {};

\fill[blue] (.2928932190,0.340741737e-1) circle (0.6pt) node {};
\fill[blue] (.2928932190,.2928932190) circle (0.6pt) node {};
\fill[blue] (.2928932190,.7411809549) circle (0.6pt) node {};
\fill[blue] (.2928932190,1.258819045) circle (0.6pt) node {};
\fill[blue] (.2928932190,1.707106781) circle (0.6pt) node {};
\fill[blue] (.2928932190,1.965925826) circle (0.6pt) node {};

\fill[blue] (.7411809549,0.340741737e-1) circle (0.6pt) node {};
\fill[blue] (.7411809549,.2928932190) circle (0.6pt) node {};
\fill[blue] (.7411809549,.7411809549) circle (0.6pt) node {};
\fill[blue] (.7411809549,1.258819045) circle (0.6pt) node {};
\fill[blue] (.7411809549,1.707106781) circle (0.6pt) node {};
\fill[blue] (.7411809549,1.965925826) circle (0.6pt) node {};

\fill[blue] (1.258819045,0.340741737e-1) circle (0.6pt) node {};
\fill[blue] (1.258819045,.2928932190) circle (0.6pt) node {};
\fill[blue] (1.258819045,.7411809549) circle (0.6pt) node {};
\fill[blue] (1.258819045,1.258819045) circle (0.6pt) node {};
\fill[blue] (1.258819045,1.707106781) circle (0.6pt) node {};
\fill[blue] (1.258819045,1.965925826) circle (0.6pt) node {};

\fill[blue] (1.707106781,0.340741737e-1) circle (0.6pt) node {};
\fill[blue] (1.707106781,.2928932190) circle (0.6pt) node {};
\fill[blue] (1.707106781,.7411809549) circle (0.6pt) node {};
\fill[blue] (1.707106781,1.258819045) circle (0.6pt) node {};
\fill[blue] (1.707106781,1.707106781) circle (0.6pt) node {};
\fill[blue] (1.707106781,1.965925826) circle (0.6pt) node {};

\fill[blue] (1.965925826,0.340741737e-1) circle (0.6pt) node {};
\fill[blue] (1.965925826,.2928932190) circle (0.6pt) node {};
\fill[blue] (1.965925826,.7411809549) circle (0.6pt) node {};
\fill[blue] (1.965925826,1.258819045) circle (0.6pt) node {};
\fill[blue] (1.965925826,1.707106781) circle (0.6pt) node {};
\fill[blue] (1.965925826,1.965925826) circle (0.6pt) node {};

%\draw[shade,draw] (2,0) rectangle (4,2);

%% Vertical lines 1
\draw [dashed] (1.889185300,0) -- (1.889185300,2);
\draw [dashed] (2.186827202,0) -- (2.186827202,2);
\draw [dashed] (2.702358098,0) -- (2.702358098,2);
\draw [dashed] (3.297641902,0) -- (3.297641902,2);
\draw [dashed] (3.813172798,0) -- (3.813172798,2);
\draw [dashed] (4.110814700,0) -- (4.110814700,2);
%% Horizontal lines 1
\draw [dashed] (2,0.340741737e-1) -- (4,0.340741737e-1);
\draw [dashed] (2,.2928932190) -- (4,.2928932190);
\draw [dashed] (2,.7411809549) -- (4,.7411809549);
\draw [dashed] (2,1.258819045) -- (4,1.258819045);
\draw [dashed] (2,1.707106781) -- (4,1.707106781);
\draw [dashed] (2,1.965925826) -- (4,1.965925826);

\fill[blue] (1.889185300,0.340741737e-1) circle (0.6pt) node {};
\fill[blue] (1.889185300,.2928932190) circle (0.6pt) node {};
\fill[blue] (1.889185300,.7411809549) circle (0.6pt) node {};
\fill[blue] (1.889185300,1.258819045) circle (0.6pt) node {};
\fill[blue] (1.889185300,1.707106781) circle (0.6pt) node {};
\fill[blue] (1.889185300,1.965925826) circle (0.6pt) node {};

\fill[blue] (2.186827202,0.340741737e-1) circle (0.6pt) node {};
\fill[blue] (2.186827202,.2928932190) circle (0.6pt) node {};
\fill[blue] (2.186827202,.7411809549) circle (0.6pt) node {};
\fill[blue] (2.186827202,1.258819045) circle (0.6pt) node {};
\fill[blue] (2.186827202,1.707106781) circle (0.6pt) node {};
\fill[blue] (2.186827202,1.965925826) circle (0.6pt) node {};

\fill[blue] (2.702358098,0.340741737e-1) circle (0.6pt) node {};
\fill[blue] (2.702358098,.2928932190) circle (0.6pt) node {};
\fill[blue] (2.702358098,.7411809549) circle (0.6pt) node {};
\fill[blue] (2.702358098,1.258819045) circle (0.6pt) node {};
\fill[blue] (2.702358098,1.707106781) circle (0.6pt) node {};
\fill[blue] (2.702358098,1.965925826) circle (0.6pt) node {};

\fill[blue] (3.297641902,0.340741737e-1) circle (0.6pt) node {};
\fill[blue] (3.297641902,.2928932190) circle (0.6pt) node {};
\fill[blue] (3.297641902,.7411809549) circle (0.6pt) node {};
\fill[blue] (3.297641902,1.258819045) circle (0.6pt) node {};
\fill[blue] (3.297641902,1.707106781) circle (0.6pt) node {};
\fill[blue] (3.297641902,1.965925826) circle (0.6pt) node {};

\fill[blue] (3.813172798,0.340741737e-1) circle (0.6pt) node {};
\fill[blue] (3.813172798,.2928932190) circle (0.6pt) node {};
\fill[blue] (3.813172798,.7411809549) circle (0.6pt) node {};
\fill[blue] (3.813172798,1.258819045) circle (0.6pt) node {};
\fill[blue] (3.813172798,1.707106781) circle (0.6pt) node {};
\fill[blue] (3.813172798,1.965925826) circle (0.6pt) node {};

\fill[blue] (4.110814700,0.340741737e-1) circle (0.6pt) node {};
\fill[blue] (4.110814700,.2928932190) circle (0.6pt) node {};
\fill[blue] (4.110814700,.7411809549) circle (0.6pt) node {};
\fill[blue] (4.110814700,1.258819045) circle (0.6pt) node {};
\fill[blue] (4.110814700,1.707106781) circle (0.6pt) node {};
\fill[blue] (4.110814700,1.965925826) circle (0.6pt) node {};

%\draw[shade,draw] (0,0) rectangle (2,0);
%% Vertical lines 1
\draw [dashed] (4.034074174,0) -- (4.034074174,2);
\draw [dashed] (4.292893219,0) -- (4.292893219,2);
\draw [dashed] (4.741180955,0) -- (4.741180955,2);
\draw [dashed] (5.258819045,0) -- (5.258819045,2);
\draw [dashed] (5.707106781,0) -- (5.707106781,2);
\draw [dashed] (5.965925826,0) -- (5.965925826,2);
%% Horizontal lines 1
\draw [dashed] (4,0.340741737e-1) -- (6,0.340741737e-1);
\draw [dashed] (4,.2928932190) -- (6,.2928932190);
\draw [dashed] (4,.7411809549) -- (6,.7411809549);
\draw [dashed] (4,1.258819045) -- (6,1.258819045);
\draw [dashed] (4,1.707106781) -- (6,1.707106781);
\draw [dashed] (4,1.965925826) -- (6,1.965925826);

\fill[blue] (4.034074174,0.340741737e-1) circle (0.6pt) node {};
\fill[blue] (4.034074174,.2928932190) circle (0.6pt) node {};
\fill[blue] (4.034074174,.7411809549) circle (0.6pt) node {};
\fill[blue] (4.034074174,1.258819045) circle (0.6pt) node {};
\fill[blue] (4.034074174,1.707106781) circle (0.6pt) node {};
\fill[blue] (4.034074174,1.965925826) circle (0.6pt) node {};

\fill[blue] (4.292893219,0.340741737e-1) circle (0.6pt) node {};
\fill[blue] (4.292893219,.2928932190) circle (0.6pt) node {};
\fill[blue] (4.292893219,.7411809549) circle (0.6pt) node {};
\fill[blue] (4.292893219,1.258819045) circle (0.6pt) node {};
\fill[blue] (4.292893219,1.707106781) circle (0.6pt) node {};
\fill[blue] (4.292893219,1.965925826) circle (0.6pt) node {};

\fill[blue] (4.741180955,0.340741737e-1) circle (0.6pt) node {};
\fill[blue] (4.741180955,.2928932190) circle (0.6pt) node {};
\fill[blue] (4.741180955,.7411809549) circle (0.6pt) node {};
\fill[blue] (4.741180955,1.258819045) circle (0.6pt) node {};
\fill[blue] (4.741180955,1.707106781) circle (0.6pt) node {};
\fill[blue] (4.741180955,1.965925826) circle (0.6pt) node {};

\fill[blue] (5.258819045,0.340741737e-1) circle (0.6pt) node {};
\fill[blue] (5.258819045,.2928932190) circle (0.6pt) node {};
\fill[blue] (5.258819045,.7411809549) circle (0.6pt) node {};
\fill[blue] (5.258819045,1.258819045) circle (0.6pt) node {};
\fill[blue] (5.258819045,1.707106781) circle (0.6pt) node {};
\fill[blue] (5.258819045,1.965925826) circle (0.6pt) node {};

\fill[blue] (5.707106781,0.340741737e-1) circle (0.6pt) node {};
\fill[blue] (5.707106781,.2928932190) circle (0.6pt) node {};
\fill[blue] (5.707106781,.7411809549) circle (0.6pt) node {};
\fill[blue] (5.707106781,1.258819045) circle (0.6pt) node {};
\fill[blue] (5.707106781,1.707106781) circle (0.6pt) node {};
\fill[blue] (5.707106781,1.965925826) circle (0.6pt) node {};

\fill[blue] (5.965925826,0.340741737e-1) circle (0.6pt) node {};
\fill[blue] (5.965925826,.2928932190) circle (0.6pt) node {};
\fill[blue] (5.965925826,.7411809549) circle (0.6pt) node {};
\fill[blue] (5.965925826,1.258819045) circle (0.6pt) node {};
\fill[blue] (5.965925826,1.707106781) circle (0.6pt) node {};
\fill[blue] (5.965925826,1.965925826) circle (0.6pt) node {};

\draw[line width=0.5mm] (0,0) -- (6,0) -- (6,2) -- (0,2) -- (0,0);
\draw[line width=0.5mm] (2,0) -- (2,2);
\draw[line width=0.5mm] (4,0) -- (4,2);
\draw[line width=0.5mm] (1.85,0) -- (1.85,2);
\draw[line width=0.5mm] (4.15,0) -- (4.15,2);

\fill[red] (1.85,0.340741737e-1) circle (0.6pt) node {};
\fill[red] (1.85,.2928932190) circle (0.6pt) node {};
\fill[red] (1.85,.7411809549) circle (0.6pt) node {};
\fill[red] (1.85,1.258819045) circle (0.6pt) node {};
\fill[red] (1.85,1.707106781) circle (0.6pt) node {};
\fill[red] (1.85,1.965925826) circle (0.6pt) node {};

\fill[red] (2,0.340741737e-1) circle (0.6pt) node {};
\fill[red] (2,.2928932190) circle (0.6pt) node {};
\fill[red] (2,.7411809549) circle (0.6pt) node {};
\fill[red] (2,1.258819045) circle (0.6pt) node {};
\fill[red] (2,1.707106781) circle (0.6pt) node {};
\fill[red] (2,1.965925826) circle (0.6pt) node {};

\fill[red] (4,0.340741737e-1) circle (0.6pt) node {};
\fill[red] (4,.2928932190) circle (0.6pt) node {};
\fill[red] (4,.7411809549) circle (0.6pt) node {};
\fill[red] (4,1.258819045) circle (0.6pt) node {};
\fill[red] (4,1.707106781) circle (0.6pt) node {};
\fill[red] (4,1.965925826) circle (0.6pt) node {};

\fill[red] (4.15,0.340741737e-1) circle (0.6pt) node {};
\fill[red] (4.15,.2928932190) circle (0.6pt) node {};
\fill[red] (4.15,.7411809549) circle (0.6pt) node {};
\fill[red] (4.15,1.258819045) circle (0.6pt) node {};
\fill[red] (4.15,1.707106781) circle (0.6pt) node {};
\fill[red] (4.15,1.965925826) circle (0.6pt) node {};
\end{tikzpicture}
\end{center}
\caption{GWRM subdomains with overlapping region, distance exaggerated  (\textit{red dots}), Chebyshev roots $x_{k}=\cos(({2k-1})\pi/{2K}),\quad k=1{\ldots}K$ (\textit{blue dots}), and the subdomain boundaries (\textit{solid black lines}).} 
\label{fig:F1}
\end{figure}

Subdomains, both in space and time, have the potential to remedy the efficiency problem. Concentrating here on spatial subdomains, it may be noted that a mere division of the spatial domain into subdomains does not have a radical influence on efficiency. Clearly, the possibility to adjust the subdomain length according to the physical terrain would optimize the solution procedure, but essentially the same global number of coefficient equations would need to be solved simultaneously. The Common Boundary-Condition method (CBC-GWRM) described in the following does, however, reduce the number of simultaneous global equations to be solved. The entire computational domain $D=\lbrace(t,x):[0,T],[0,L]\rbrace$ is discretized into $s$ overlapping spatial elements, see Figure~\ref{fig:F1}. For example $s=3$ would give a discretized domain $\Omega=\lbrace\Omega_1=[0,x_1+\varepsilon],\Omega_{2}=[x_1-\varepsilon,x_{2}+\varepsilon],\Omega_{3}=[x_{2}-\varepsilon,L]\rbrace$, where $\varepsilon$ is a small overlapping distance. This procedure allows us to use only a few Chebyshev modes in our ansatz for each subdomain. By overlapping the subdomains, point-wise and gradient continuity is ensured.
The system of equations we wish to solve include $N_s$ spatial subdomains, $N_e$ number of physical equations, $K$ temporal modes, $L$ spatial modes (we here let $\Lambda=0$). This results in $N=N_sN_e(K+1)(L+1)$ equations to solve for the coefficients of the ansatz \ref{eq:M2}. A question arises: can the global amount of equations be reduced in some way?

\begin{figure}[H]
\centering
{\includegraphics[height=2.3in,width=1.0\textwidth]{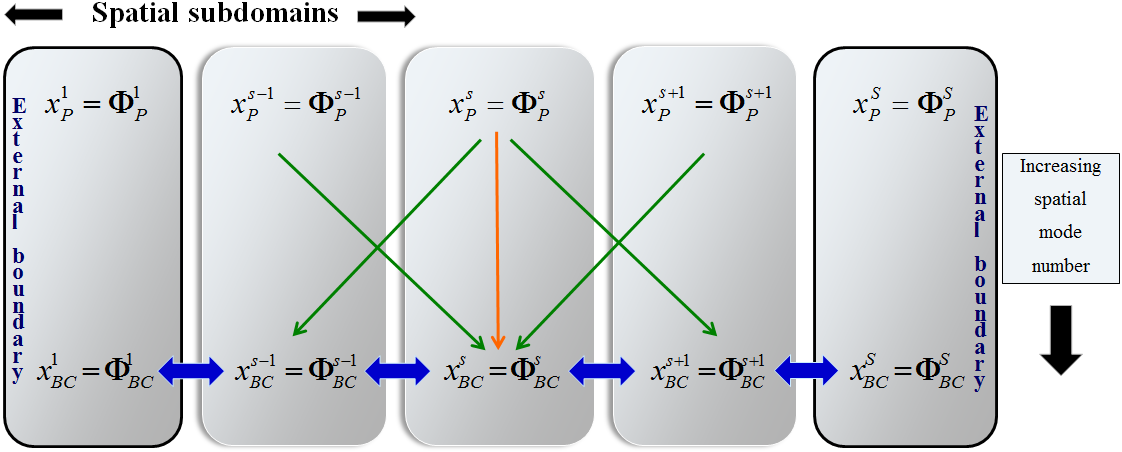}}
\caption{CBC subdomain schematic with algebraic equations indicated. The private variables (index "p") are explicitly dependent on the boundary variables (index "BC") in the given subdomain and implicitly on those of the neighbouring subdomains.} 
\label{fig:F2}
\end{figure}

Equations (6) can be written in fixed point form formally as $\textbf{x}^s=\mathbf{\Phi}^s$, where $s$ refers to the respective subdomain; for details see \cite{Scheffel:GWRM2}. The idea now is to realize that only the subset $\textbf{x}_{BC}^s=\mathbf{\Phi}_{BC}^s$ representing the boundary equations need to be solved globally. The reason for this is that the physics equations (6) in each individual subdomain $\textbf{x}_{P}^s=\mathbf{\Phi}_{P}^s$, representing the systems of PDEs, are only dependent on $\textbf{x}_{BC}^s$. These "private" equations can be solved locally in each spatial subdomain in each iteration level. This process can be fully parallellized. Furthermore, we have the dependence $\textbf{x}_{BC}^s=\mathbf{\Phi}_{BC}^s(s-1,s,s+1)$, depicted graphically in Figure \ref{fig:F2}. Thus boundary condition equations only contain Chebyshev coefficients (variables) of the immediately neighbouring spatial subdomains.

A numerical example may be elucidating. Let us assume that a system of five PDEs, representing second order in space, is solved in 1D. Employing $K=9$ and 10 spatial subdomains with $L=9$, a total of 5000 global equations result for the Chebyshev coefficients of Eq. (2). The CBC-GWRM algorithm reduces these to 1000. For 2D and 3D problems the reduction is even more profound.
The reduced set of global equations is then solved iteratively by SIR \cite{Scheffel:SIR}. SIR solves the algebraic set of equations by iterating the following equation,
\begin{align}
\textbf{x}^{i+1}_{BC}=[\textbf{I}+(\textbf{R}^i-\textbf{I})(\textbf{J}^i_{BC})^{-1}](\textbf{x}^i_{BC}-\pmb{\varphi}^i_{BC})+\pmb{\varphi}^i_{BC}~{\equiv}~\mathbf{\Phi}^i_{BC}(\textbf{x}^i_{BC}),
\label{eq:M7}
\end{align}
where $\pmb{\varphi}$ denotes the right hand side of Eq. (6) plus the boundary equations, the matrix $\textbf{R}$ with elements $R_{mn}=\partial{\Phi_m}/\partial{x_n}$ controls convergence in SIR and $\textbf{I}$ is the identity matrix. It should be noted that Eq. (7) is a formal representation; for efficiency the matrix inversion is generally avoided and a matrix equation is solved instead. Thus the Jacobian $\textbf{J}^i_{BC}$ needs to be computed, 
\begin{align}
J_{BCpq}=\frac{{\partial}({\mathbf{x}}_{BC}-\pmb{\varphi}_{BC})_p}{{\partial}{x}_{BCq}}={\delta}_{pq}-T_{pq},
\label{eq:M8}
\end{align}
where $\delta_{pq}$ is the Kronecker delta and $T_{pq}$ includes the explicit and implicit derivatives;
\begin{align}
T_{pq}=\frac{{\partial}\varphi_{BCp}}{{\partial}x_{BCq}}+\sum^{\nu=s+1}_{\nu=s-1}\sum^{N_eN_p}_{i=1}\frac{{\partial}\varphi_{BCp}}{{\partial}x^{\nu}_{Pi}}\frac{{\partial}x^{\nu}_{Pi}}{{\partial}x^{\nu}_{BCq}}.
\label{eq:M9}
\end{align}
The indices $p$ and $q$ refer to common BC variables and the index $i$ refers to the $N_p$ private variables. The first term is the explicit derivative. The second term shows that the common BC variables are indirectly dependent on the private variables in the neighbouring subdomains, hence the implicit derivatives. The index $\nu$ is introduced in the sum to neglect subdomains that are not directly influencing the current common BC variables. 

The ${{\partial}x_{P}}/{{\partial}x_{BC}}$ coefficients in the sum require some attention. The first step is to create a vector of all private equations $f_i=z_i-\varphi_i(z)$, $i=1{\ldots}N_eN_p$, where $z$ and  $\varphi(z)$ only contain the private variables and equations from $x$ and $\varphi(x)$. Since the private $z$ variables possess an implicit dependence on the common BC variables, the partial derivatives can be computed and saved in the matrix $F_{ij}={\partial}f_i/{\partial}z_j$, $i=1{\ldots}N_eN_p$ and $j=1{\ldots}N_eN_{BC}$. The implicit derivatives in $F_{ij}$ are evaluated with current private values from current scheme iteration and the ${{\partial}x_{Pi}}/{{\partial}x_{BCj}}$ variables are obtained by solving the linear algebraic system of equations $F_{ij}=0$.

An alternative method for obtaining the ${{\partial}x_{P}}/{{\partial}x_{BC}}$ coefficients in Eq.~\ref{eq:M9} is to approximate the derivative numerically with forward difference ${{\partial}x_{P}}/{{\partial}x_{BC}}=[x_{P}(x_{BC}+\epsilon)-x_{P}(x_{BC})]/\epsilon$. The set of private equations numbering $(L-1)(K+1)$ have to be solved for every BC variable, i.e. $2N_sN_e(K+1)$ times. The amount of operations can be further reduced since the calculated Jacobian from the regular solution $x_{P}(x_{BC})$ can be used for the perturbed solution $x_{P}(x_{BC}+\epsilon)$ as well. 

%\pagebreak 
A work flow of the Common Boundary Condition subdomain scheme, employing the latter numerical scheme for the Jacobian, applied to the 1D case, omitting physical parameter dependence, is given in the box below. \\

\begin{mdframed}
\begin{center}
\textbf{CBC -  subdomain scheme (1D, $\Lambda=0$)}
\end{center} ~\\
{\small \sloppy \textbf{Comment:} Define the bounded computational domain $D=\lbrace(t,x):[t_0,t_1],[x_0,x_1]\rbrace$. Then discretize the spatial domain into overlapping subdomains $N_s$. The $N_e$ partial differential equations, with initial conditions, are then spectrally decomposed. Assume here 2 external boundary conditions, so that $L-1$ spatial modes of Eq. \ref{eq:M6} are computed.
\\ \\
\textbf{Precompute:} Total set of equations $x_j\!\!=\!\!\varphi_j(\textbf{x}),~j\!\!=\!\!1..N$, where $N=N_eN_s(K+1)(L+1)$ is the total number of equations to solve. Create set of private equations $\textbf{x}_P^s=\pmb{\varphi}^s_P(\textbf{x}^s_P,\textbf{x}_{BC}^s)$ numbering $N_P=N_eN_s(K+1)(L-1)$, and common BC equations $\textbf{x}_{BC}=\pmb{\varphi}_{BC}(\textbf{x}_{BC})$ numbering $N_{BC}=2N_eN_s(K+1)$ by extracting the appropriate indexes from the global set of equations.
\\ \\
\textbf{Step 1:} Initial common boundary condition values $\textbf{X}_{BC}^s$ are substituted into $\textbf{x}_P^s=\pmb{\varphi}^s_P(\textbf{x}^s_P,\textbf{X}_{BC}^s)$, so that only private variables are unknown. Call on SIR to solve private equations. 
\\ \\
\textbf{Step 2:} Compute ${{\partial}x_{P}}/{{\partial}x_{BC}}\!\!\!=\![x_{P}(x_{BC}\!+\epsilon)\!-x_{P}(x_{BC})]/\epsilon$ by solving $\textbf{x}^s_P=\pmb{\varphi}^s_P(\textbf{x}^s_P,\textbf{X}^s_{BC})$, for every BC variable $X_{BC}=X_{BC}+\varepsilon$, i.e. $2N_sN_e(K+1)$ times. Here $\varepsilon$ is the finite difference length.
\\ \\
\textbf{Step 3:} Compute the BC Jacobian (Eqs.~\ref{eq:M8}-\ref{eq:M9})
\begin{align*}
J_{BCpq}={\delta}_{pq}-\frac{{\partial}\varphi_{BCp}}{{\partial}x_{BCq}}-\sum^{\nu=s+1}_{\nu=s-1}\sum^{N_eN_p}_{i=1}\frac{{\partial}\varphi_{BCp}}{{\partial}x^{\nu}_{Pi}}\frac{{\partial}x^{\nu}_{Pi}}{{\partial}x^{\nu}_{BCq}}.
\end{align*}
\textbf{Step 4:} Compute Eq.~\ref{eq:M7}, $\textbf{x}^{i+1}_{BC}=\mathbf{\Phi}^i_{BC}(\textbf{x}^i_{BC})$, which includes $\textbf{J}_{BC}$ from step 3) and $\pmb{\varphi}_{BC}(\textbf{x}_{BC})$ with private variables from step 1). This here is a linear system, hence it only requires a single SIR iteration to calculate the new common BC values  $\textbf{x}_{BC}^{i+1}$.
\\ \\
\textbf{Step 5:} Repeat step 1), i.e. compute also $\textbf{x}_P^s=\pmb{\varphi}^s_P(\textbf{x}^s_P,\textbf{X}_{BC}^s)$, where $\textbf{X}_{BC}=\textbf{x}_{BC}^{i+1}$ from step 4).
\\ \\
\textbf{Final:} Update Chebyshev coefficients $a_{kl}$ in Eq. \ref{eq:M2} using solution vector $\textbf{x}$.}
\end{mdframed}
~\\

\section{Test problems}
We have chosen to solve the linearised Burger equation, a forced wave equation, and a problem modelled with the linearised ideal MHD equations. Non-linear problems will be addressed in a forthcoming paper. All CBC-GWRM simulations are carried out on one global temporal interval, that is we here omit the use of time intervals.

\subsection{Linearised 1D Burger equation}
\sloppy The performance of the CBC-GWRM with regards to accuracy is here compared to that of the explicit Lax-Wendroff method, the implicit Crank-Nicolson method, and the "standard" GWRM with a global spatial subdomain scheme. The purpose of employing the Crank-Nicolson and Lax-Wendroff methods is to include two well known finite difference methods that require little justification. The performance of other, more or less sophisticated, finite difference methods can easily be compared to these methods.

The parameters that determine the accuracy and efficiency of the GWRM are the order of temporal Chebyshev modes $K$, spatial Chebyshev modes $L$, and the number of spatial subdomains $N_s$. Similarly, the grid values determine the performance of the FD methods, namely $M$ temporal steps and $N$ spatial steps. The CPU time and memory consumption of all the methods are documented throughout.

\begin{figure}[H]
\centering
\subfloat[Solution]{\includegraphics[width=0.48\textwidth]{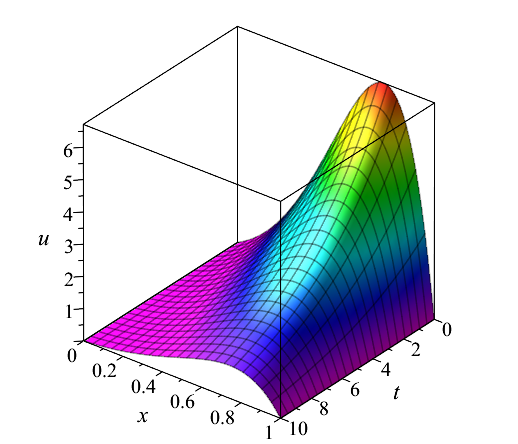}\label{fig:B1a}}
  \hfill 
  \subfloat[Error]{\includegraphics[width=0.52\textwidth]{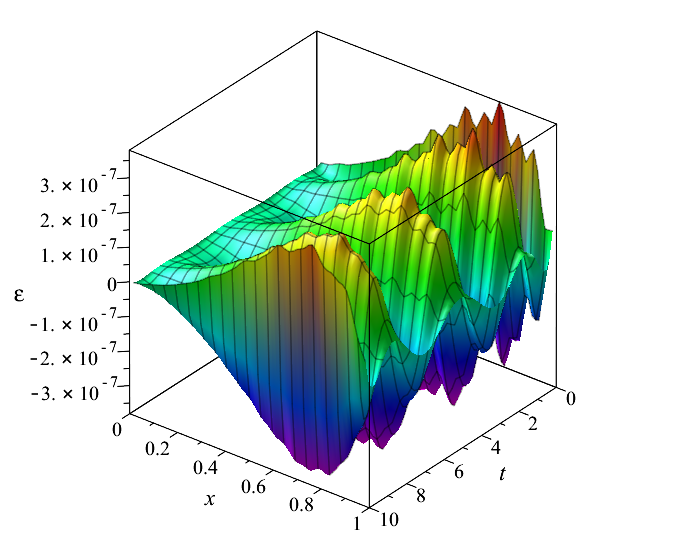}\label{fig:B1b}}
  \caption{CBC-GWRM solution and error with $\kappa=0.01$ and $b=0.06$; $K=7$, $L=6$ and $N_s=7$.}
%\label{fig:B1}}
\end{figure}

The linearised 1D Burger equation is stated as follows:
\begin{align}
\frac{{\partial}u}{{\partial}t}+b\frac{{\partial}u}{{\partial}x}={\kappa}\frac{{\partial}^2u}{{\partial}x^2}.
\label{eq:diff}
\end{align}
\noindent
where $\kappa$ and \textit{b} are constants. The initial condition and boundary conditions chosen here are
\begin{align}
u(0,x)=\text{sin}({\pi}x)e^{{bx}/{2{\kappa}}}, \\
u(t,0)=u(t,1)=0.
\label{eq:ini_diff}
\end{align}
This enables an exact solution that can be used for accurate benchmarking (see Figures \ref{fig:B1a} and \ref{fig:B1b} for an example),
\begin{align}
u(t,x)=\text{sin}({\pi}x)e^{{bx}/{2{\kappa}-(b^2/4\kappa+\kappa\pi^2)t}}
\label{eq:ex_diff}
\end{align}
The CPU time and memory consumption of the CBC-GWRM have been compared with two finite difference methods. These are the explicit Lax-Wendroff and the implicit Crank-Nicolson methods. For comparable accuracies of $\varepsilon\sim{10^{-3}}$ the CBC-GWRM features a CPU runtime of $0.16~s$ and a memory consumption of $20~\text{MB}$, using a Maple implementation on a desktop PC. This was achieved with the parameters $N_s=2$, $K=4$, and $L=5$. The Lax-Wendroff method with $M=340$ time steps and $N=45$ spatial steps featured a runtime of $0.94~s$ and memory $54~\text{MB}$. The Crank-Nicolson method required $0.09~s$ and $20~\text{MB}$ with parameters $M=60$ and $N=80$. Thus we see that for moderate accuracies the C-N method is approximately two times faster than CBC-GWRM and 10 times faster than L-W, all methods requiring comparable memory consumption. 

However, when the accuracy is increased to $\varepsilon\sim{10^{-5}}$, the CBC-GWRM with $N_s=3$, $K=6$, and $L=6$ features a CPU time of $0.50~s$ and memory of $22~\text{MB}$. The C-N method gave a CPU time of $0.77~s$ and memory of $43~\text{MB}$ for $M=120$ and $N=400$, whilst the L-W method was unable to reach a comparable accuracy. Thus the CBC-GWRM is here faster and more memory efficient for higher accuracies than both FD-methods. With the same parameters and accuracy $\varepsilon\sim{10^{-5}}$ the "standard" GWRM (without the CBC-scheme) achieves a CPU time $t=0.27~s$ and memory $23~\text{MB}$. In Discussion, however, it is argued that the CPU time of the GWRM using the CBC scheme is strongly advantageous when a large number of spatial subdomains is employed.
\begin{figure}[H]
\center
\includegraphics[height=2.5in,width=2.5in]{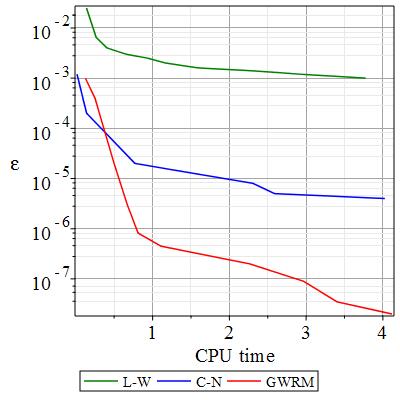}
\caption{Error plot vs CPU time [s]} 
\label{fig:B3}
\end{figure}
The three methods have been analysed with optimal parameters so as to obtain the best accuracy for a given computational time. The data is displayed in Figure \ref{fig:B3}, which shows how all three methods scale in this regard. A simple curve-fit of the data shows that the L-W method scales as $\varepsilon_{LW}~{\sim}~t_{CPU}^{-0.6}$, and the C-N method scales slightly better at $\varepsilon_{CN}~{\sim}~t_{CPU}^{-0.9}$. The spectral convergence properties of the CBC-GWRM allows a much stronger scaling than the FD methods; $\varepsilon_{GWRM}~{\sim}~t_{CPU}^{-3.47}$. It is also found that the CPU time scales linearly as $t_{CPU}~{\sim}~N_s^1$ with regards to the number of subdomains used, see Figure \ref{fig:D2}, in Discussion. This feature of the CBC-GWRM becomes advantageous when modelling physical systems that require high local spatial accuracy. 

\subsection{Forced Wave Equation}
\begin{figure}[H]
\begin{center}
\makebox[0pt]{\includegraphics[height=1.8in,width=0.9\textwidth]{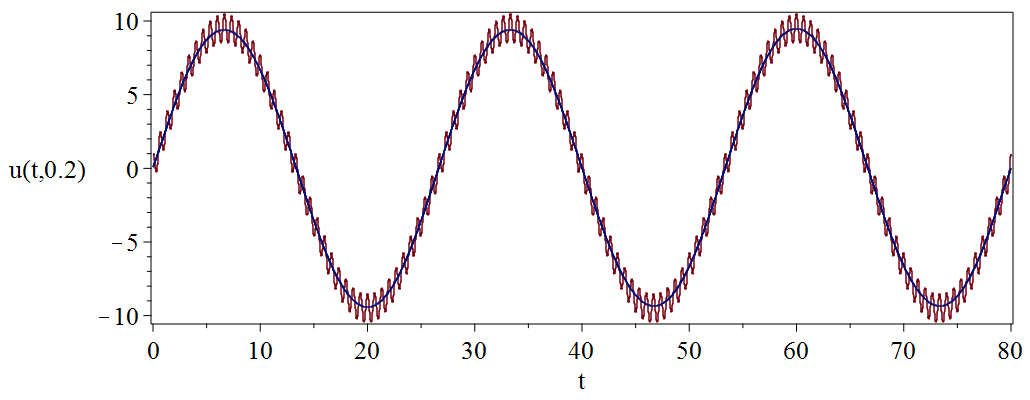}}
\end{center}
\caption{Temporal plot for interval $[0,80]$ at $x=0.2$ showing how the CBC-GWRM (\textit{blue line}) resolves the slower time scale of the exact solution (\textit{red line}). The parameters used are $N_s=4$, $K=14$, and $L=5$.} 
\label{fig:W2}
\end{figure}

\begin{figure}[H]
\centering
  \subfloat[CPU times]{\includegraphics[width=0.50\textwidth]{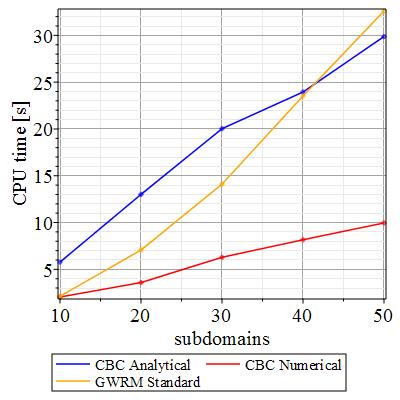}\label{fig:WE2}}
  \hfill 
  \subfloat[Memory consumption]{\includegraphics[width=0.50\textwidth]{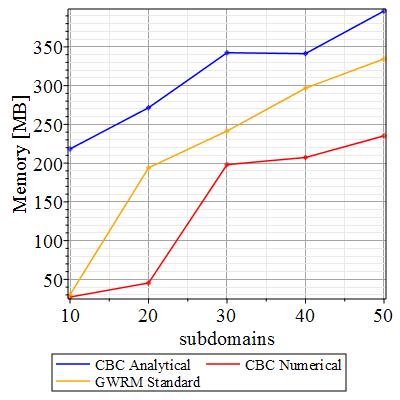}\label{fig:WE3}}
  \caption{Comparison between the standard (global) GWRM and the CBC-scheme for the wave equation, with analytical derivatives and numerical derivatives. The subdomain scaling (a) and memory consumption (b) is plotted for parameters $A=10$, $n=3$, $T=30$, $\alpha=2\pi/T$, $\beta=3\pi$ and $c=1$ with time interval $t=[0,30]$.}
\end{figure}

The forced wave equation employed here is a second order hyperbolic differential equation that features two time scales. This equation is used to test efficiency, in the sense that in some cases accuracy at small scales may be ignored in favour of efficiently resolving large scale dynamics. The wave equation has the form
\begin{align}
\frac{{\partial}^2u}{{\partial}t^2}=c^2\frac{{\partial}^2u}{{\partial}x^2}+f(t,x).
\label{eq:w_e}
\end{align}
and can be posed as two first order in time partial differential equations:
\begin{align*}
\frac{{\partial}v}{{\partial}t}=c^2\frac{{\partial}^2u}{{\partial}x^2}+f(t,x), \\
\frac{{\partial}u}{{\partial}t}=v.
\end{align*}
The initial and boundary conditions are
\begin{align*}
u(0,x)=\text{sin}({\pi}x), \\
v(0,x)={\alpha}A\text{sin}({\beta}x), \\
u(t,0)=u(t,1)=0,  \\
v(t,0)=v(t,1)=0.
\end{align*}
Here $A$, $n$, $\alpha$, $\beta$ and $c$ are free parameters.  The forcing equation chosen is
\begin{align*}
f(t,x)=A(c^2{\beta}^2-{\alpha}^2)\text{sin}({\alpha}t)\text{sin}({\beta}x).
\end{align*}
This gives the exact solution
\begin{align}
u(t,x)=\text{cos}(nc{\pi}t)\text{sin}(n{\pi}x)+A\text{sin}({\alpha}t)\text{sin}({\beta}x).
\label{eq:ex_ew}
\end{align}
First the free parameters are set to $A=10$, $n=3$, $T=80$, $\alpha=6\pi/T$, $\beta=3\pi$ and $c=1$. For the time interval $t\in[0,80]$ the CBC-GWRM requires a high number of temporal Chebyshev modes $K{\geq}12$ to achieve a reasonable average (within 92\% percent of the second, slowly evolving part of the solution (15)) of the fast time scale dynamics. The solution at $x=0.2$ can be seen in Figure \ref{fig:W2}.
Since it is inefficient to solve for more than one wavelength with high order temporal modes, a more pragmatic test would be to solve the wave equation with parameters $A=10$, $n=3$, $T=30$, $\alpha=2\pi/T$, $\beta=3\pi$ and $c=1$ in a time interval $t\in[0,30]$. This allows a temporal mode $K=5$ to accurately average the fast time scale (within 95\% percent of the slow manifold). The CBC-GWRM with parameters $K=5$, $L=5$, and $N_s=4$ achieves a CPU time of $0.86~s$, which is comparable to the standard GWRM subdomain scheme with the same parameters, requiring $0.83~s$. In Figure~\ref{fig:WE2} the standard GWRM and the CBC-scheme, employing both analytical and numerical derivatives, for difference numbers of increasing subdomains. The CBC-scheme with numerical derivatives  achieves shorter CPU times and is more memory efficient. In Figure~\ref{fig:WE3} the memory consumption of the three GWRM schemes are compared. 
\iffalse
\begin{figure}[H]
\begin{center}
\makebox[0pt]{\includegraphics[height=2.8in,width=1.1\textwidth]{Wave_Large.png}}
\end{center}
\caption{{\color{blue}CBC-GWRM $u(t,x)$ (\textit{left}) and exact solution $u_e(t,x)$ (\textit{right})} of the forced wave equation in the range $t=[0,80]$. CBC-GWRM parameters are $N_s=5$, $K=12$, and $L=5$.}
\label{fig:W1}
\end{figure}
\fi
Turning to the finite difference methods, the L-W method is able to accurately resolve the slow time scale with parameters $M=1200$ and $N=40$ in $1.5~s$. The L-W method initially follows the fast time scale until the solution advances further away from the initial condition, in which case it transitions to the slow time scale. 

The parameters used for the C-N solution is $M=50$ and $N=50$, which gives a run time of $1.4~s$. The C-N method fails to average correctly as it is slightly out of phase. Explicit methods, unlike implicit methods, are conditionally stable because they must obey the CFL condition $\gamma{\vartriangle}t/{\vartriangle}x{\leq}1$, where $\gamma$ is the wave speed, ${\vartriangle}t$ is the temporal step length, and ${\vartriangle}x$ is the spatial step length. However, since the forced wave equation is a hyperbolic equation, the FD methods need to resolve the phase of the wave in order to be accurate. Furthermore, the phase can only be resolved by following the strong stability criterion, in which case the implicit method shows no favourable qualities over explicit schemes \cite{Appadu:1}.

To conclude, the CBC-scheme is approximately $1.6$ times faster than the finite difference methods when one wavelength is simulated with a $K=5$ temporal Chebyshev mode. Similar, and even higher, accuracy and efficiency levels can be attained using lower temporal modes and multiple time intervals than for high temporal modes with few time intervals. The memory consumption of the CBC-GWRM is also superior in this regard compared to the GWRM global subdomain scheme. It can be seen in Figure~\ref{fig:WE3}; if a low temporal mode $K=5$ and about $20$ subdomains are used the CBC-GWRM is 75\% more memory efficient and 50\% faster than the global GWRM. These are substantial increases in performance which allows the CBC-GWRM to be highly competitive when solving physical systems with high degrees of freedom.

\subsection{Ideal MHD}
The ideal MHD equations are included in our analysis to investigate whether the CBC-GWRM is capable of accurately computing complex physical systems. The ideal MHD model is a set of coupled partial differential equations that describe the dynamics of a perfectly conductive fluid. The ideal MHD equations provide a simple description of plasma dynamics. The equations are stated as
\begin{align}
\label{cont.eq}
\frac{{\partial}\rho}{{\partial}t}+\nabla\cdot(\rho\textbf{v})=0, \\
\label{motion.eq} 
\rho\Big(\frac{{\partial}}{{\partial}t}+\textbf{v}\cdot\nabla\Big)\textbf{v}+\nabla{p}-\textbf{J}\times\textbf{B}=0, \\
\label{energy.eq}
\frac{{\partial}p}{{\partial}t}+\nabla\cdot(p\textbf{v})+(\Gamma-1)p\nabla\cdot\textbf{v}=0,  \\
\label{Faraday.eq}
\frac{{\partial}\textbf{B}}{{\partial}t}+\nabla\times\textbf{E}=0, \\
\label{Ampere.eq}
\nabla\times\textbf{B}-\mu_0\textbf{J}=0, \\
\label{Ohm.eq}
\textbf{E}+\textbf{v}\times\textbf{B}=0. 
\end{align}
First we have the fluid equations; the continuity Eq.~\ref{cont.eq} that describes the conservation of mass density $\rho$; the equation of motion (\ref{motion.eq}) which solves for the fluid velocity $\textbf{v}$; and the energy equation (\ref{energy.eq}) describing the evolution of the pressure profile $p$;
\begin{align}
\label{energy2.eq}
\Big(\frac{{\partial}}{{\partial}t}+\textbf{v}\cdot\nabla\Big)\Big(\frac{p}{\rho^\Gamma}\Big)=0,
\end{align}
where $\Gamma=5/3$ is the ratio of specific heats. The second set involves the electromagnetic equations that describe the evolution of the magnetic field $\textbf{B}$, electric field $\textbf{E}$, and the current density $\textbf{J}$; Faraday's law (\ref{Faraday.eq}), Ampere's law (\ref{Ampere.eq}), where $\mu_0$ is the permeability in vacuum, and Ohm's law (\ref{Ohm.eq}). 

The goal here is to solve for ideal MHD instabilities that occur in a magnetically confined cylindrical plasma with coordinates $(r,\theta,z)$. The CBC-GWRM is applied to the linearised ideal MHD equations about an equilibrium, which consequently consists of 14 (7 real and 7 imaginary) coupled partial differential equations. A perturbation ${\varpropto}~\text{exp}[i(m\theta+kz)]$ is introduced, where $m$ and $k$ are the azimuthal and transverse mode numbers, respectively. 

The inner boundary conditions need to be handled with some care to avoid singularities at $r=0$ \cite{Lundin:1}. The outer boundary $r=1$ condition is that of a perfectly conducting wall, hence the radial variable components are set to zero at the wall.
\begin{figure}[H]
\centering
  \subfloat[Perturbed radial velocity]{\includegraphics[width=0.49\textwidth]{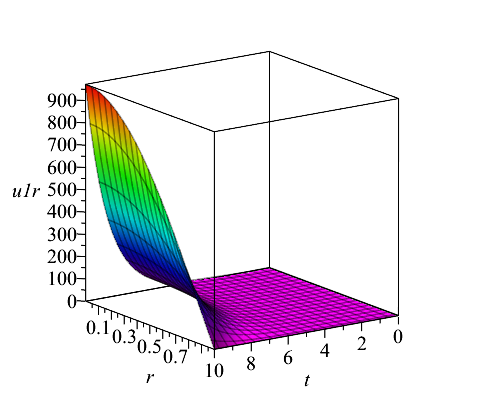}\label{fig:f41a}}
  \hfill 
  \subfloat[Perturbed radial pressure]{\includegraphics[width=0.50\textwidth]{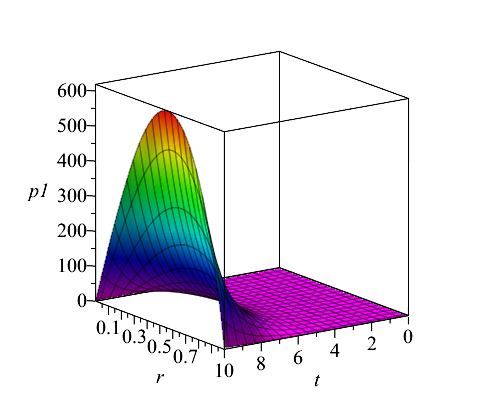}\label{fig:f41b}}
  \caption{Perturbed screw-pinch, CBC-GWRM $K=7$, $L=5$ and $N_s=3$.}
\end{figure}
The equilibrium chosen is a simple screw-pinch equilibrium (in normalized variables),
\begin{align*}
B_r=0, ~ B_t=r, ~ B_z=0.05, \\
p_0=1-r^2, ~ J_t=0, ~ J_z=2.
\end{align*}
The equilibrium was then perturbed using $m=1$ and $k=5$. The time evolution of the perturbed radial velocity and pressure is shown in Figures \ref{fig:f41a} and \ref{fig:f41b}. Initially the perturbed variables are dominated by a host of different waves, which is why the simulation has to advance far enough for the dominating unstable mode to become distinguishable. 
\begin{figure}[H]
\centering
{\includegraphics[width=0.5\textwidth]{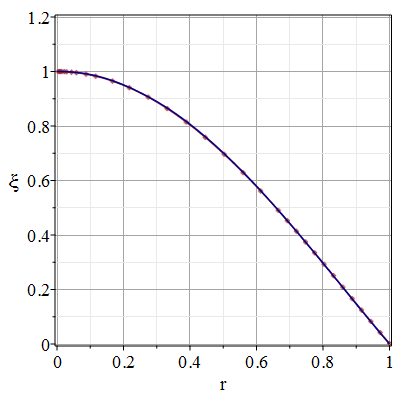}\label{fig:f51}}
\caption{Eigenfunction $\xi(r)$ of perturbation $m=1$, $k=5$. CBC-GWRM (\textit{line}) mode growth rate $\omega_{CBC}=0.839$, $K=7$, $L=5$ and $N_s=3$; Bateman $\omega_B=0.840$ (\textit{point-line})}
\end{figure}
The growth rate obtained from the CBC-GWRM is compared to that of a shooting code developed in \cite{Bateman:1}. The CBC-GWRM calculated a growth rate $\omega_{CBC}=0.839$ whilst the shooting code obtained a growth rate $\omega_B=0.840$. For higher accuracy it is advised to use multiple time intervals with fewer temporal modes. Also, this linearised MHD test shows the advantage of using a Chebyshev spectral ansatz because of its ability to average over small scale dynamics. The initial waves that propagate are of no interest, so they are averaged out, which leaves the dominating unstable mode.

\section{Discussion}
The CBC-GWRM has several favourable properties; 1) rapid  spectral convergence and high accuracy is achieved in both time and space, 2) it uses real-valued Chebyshev polynomials having the useful mini-max property, 3) sparse matrix methods can be used for solving the coefficient equations, 4) the use of reduced global coefficient matrix equations where only the internal and external boundary conditions need be solved, and 5) the subdomain private equations can be parallellized. The first three points are common properties of time-spectral methods, while this work is focused on points 4 and 5. Regarding 4), in the example problems of this paper the number of operations is reduced by a factor $\sim(2/L)^3$ and the memory usage by $\sim(2/L)^2$ since only the internal and external boundary conditions are solved globally. The "2" comes from the fact that only the two highest spatial Chebyshev modes are allocated for the two boundary conditions employed. 

It was found that the CBC-GWRM gained quite a substantial amount of computational speedup when the ${\partial}x_P/{\partial}x_{\text{BC}}$ variables were computed numerically rather than analytically. This is shown in Figure~\ref{fig:WE2}. Either forward difference or center difference can be employed. It is recommended to use forward difference for optimal efficiency, whilst achieving similar accuracy.

\begin{figure}[H]
\centering
\includegraphics[height=2.4in,width=0.5\textwidth]{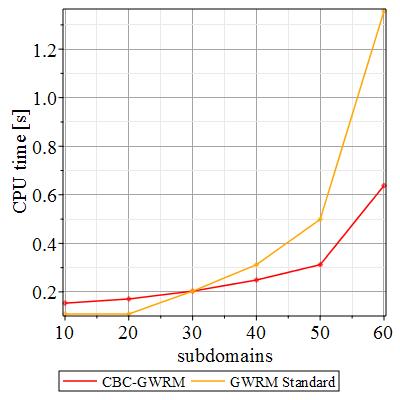}
\caption{Total CPU time [s] for solving all corresponding algebraic equations with increasing subdomains for 1D linearised Burger computation; CBC-GWRM (\textit{red}) and standard GWRM (\textit{yellow}), with the same parameters $K=5$ and $L=5$.}
\label{fig:D1}
\end{figure}

\begin{figure}[H]
\centering
\includegraphics[height=2.6in,width=0.50\textwidth]{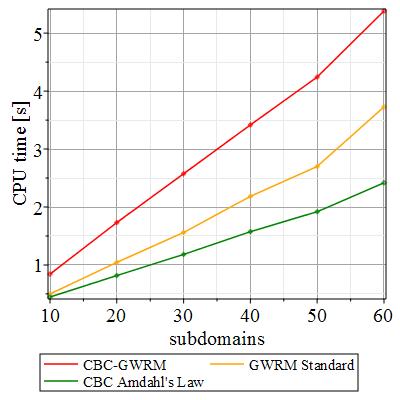}
\caption{CPU time [s] with increasing subdomains for 1D linearised Burger computation; CBC-GWRM (\textit{red}), standard GWRM (\textit{yellow}), and an approximated parallelized time that follows Amdahl's law for CBC-GWRM (\textit{green}). All methods use parameters $K=5$ and $L=5$.}
\label{fig:D2}
\end{figure}

We can find a scaling law estimate for the efficiency of the CBC-GWRM. The number of private equations for one physical equation in each subdomain of, for example the Burger equation is $(L-1)(K+1)$. The private set of equations need to be solved twice for each common boundary condition and temporal mode. The total number of operations is then $2N_s(K+1)(L-1)^3(K+1)^3$ when a single PDE is solved. The number of global common boundary condition equations are $2N_s(K+1)$. The standard GWRM scheme with sparse band matrix algorithms scales approximately as $N_s^{1.4}$. The same procedure is used to solve the BC equations in the CBC-method. Thus the ratio of the total number of operations for the GWRM-CBC and GWRM methods is 
\begin{align}
\frac{2N_s(L-1)^3(K+1)^4+N_s^{1.4}2^3(K+1)^3}{N_s^{1.4}(L+1)^3(K+1)^3}=\frac{2(K+1)}{N_s^{0.4}}\Big(\frac{L-1}{L+1}\Big)^3+\frac{2^3}{(L+1)^3}.
\label{eq:23}
\end{align}
From Eq.~\ref{eq:23} it can thus be seen that the CBC-GWRM is more efficient than the standard GWRM when $N_s$ is large. For example; $L=5$ and $K=5$ gives the criterion that $N_s{\gtrsim}27$ for the CBC-GWRM to be more efficient. This is an important result since many fluid dynamics applications suggest $N_s>100$. The scaling law Eq.~\ref{eq:23} only takes into account the amount of operations it takes to solve the total amount of algebraic equations for both the CBC-GWRM and the standard GWRM, i.e. \textit{excluding all overhead operations}, such as calculating the Jacobian matrices; see Figure~\ref{fig:D1} to see actual CPU times consistent with the scaling argument in Eq.~\ref{eq:23}. The effect of the first term in Eq.~\ref{eq:23} can be reduced substantially, employing parallellization. Thus, the limiting dependence $(2/(L+1))^3$ can be approached.
It is not surprising that the unoptimized CBC-GWRM, without parallel implementation, is slower in some cases than its global GWRM counterpart since the CBC-scheme requires extra overhead operations. It should be noted that the CBC-scheme will outperform the global GWRM in all cases once a sufficient amount of subdomains is used, parallelized or not (see Figure \ref{fig:WE2}). The work done in each subdomain can be allocated to multiple processors, which would decrease the computation time. Figure \ref{fig:D2} shows how the CPU time, \textit{including overhead operations}, for the linearised Burger equation, scales with the number of subdomains $N_s{\in}[10,60]$ for the CBC-GWRM, the standard GWRM with all equations solved globally, and the CBC-GWRM adjusted for approximate parallelizable speedup gains. The standard GWRM outperforms the CBC-GWRM in this simple case. However, when approximations of the parallel speed-up gains are made we see that CBC-GWRM could potentially outperform the standard GWRM, whilst featuring less memory consumption. The estimation of the parallelizable time was made from the speed-up gains ascertained from Amdahl's Law \cite{AFIPS97}. Amdahl's law is formulated as $S=1/[(1-p)+p/s]$, where $S$ is the speed-up gain, $s$ is the proportion of the code that can be computed in parallel,  and $p$ is the unoptimized computation time of $p$. If parallelism is accounted for the difference in CPU time can be resolved and improved upon (see Figure \ref{fig:D2}). 
The avenue of parallelizable subdomains grows even more advantageous when studying the ideal MHD equations. For the ideal MHD case approximately 95\% of the code can be parallelized, which according to Almdahl's Law gives a speedup $>12$ if more than 32 processors are used and it saturates at $\text{speedup}=20$ for cores numbering 2048 and higher \cite{AFIPS97}.

To close, it has been widely held that time-spectral methods are not as effecient as their finite difference counterparts. While spectral methods do feature higher accuracies, the idea of lacking efficiency has been challenged by such methods as the GWRM and the spectral deferred correction method. The spectral deferred correction method still, most commonly, relies on implicit and explicit finite difference methods to create a crude initial approximation, which is then corrected. This is achieved by using spectral integration in-between the finite time steps which is included to correct the crude approximation. This makes the spectral deferred correction method a high order method, requiring longer computation times \cite{Gustafsson:2,Gustafsson:1,Speck:1}.

\section{Conclusion}
The CBC-GWRM solves a set of initial-value ordinary or partial differential equations in the temporal and spatial domains with a time-spectral weighted residual method. This allows the method to obtain high accuracy and to efficiently average over small scale dynamics, as seen in the modelling of the linearised Burger and forced wave equations. Before this work the GWRM solved all subdomains simultaneously from the global set of algebraic equations. Our goal was to break down the problem into smaller pieces for enhancing efficiency. The CBC-method provides a solution to this problem.

For the 1D linearised Burger equation the CBC-GWRM was compared to finite difference methods. At low accuracy the implicit finite difference scheme outperformed both the explicit and the time-spectral methods. The maximum error, however, of the CBC-GWRM solution scales as $\varepsilon_{GWRM}~{\sim}~t_{CPU}^{-3.47}$, compared to $\varepsilon_{LW}~{\sim}~t_{CPU}^{-0.6}$ and $\varepsilon_{CN}~{\sim}~t_{CPU}^{-0.9}$. This allows CBC-GWRM to be 30\%  faster than Crank-Nicolson for accuracies $\varepsilon{\sim}10^{-5}$. The approximated parallellized CBC-GWRM CPU time also scales as $N_s^{1.1}$, with increasing number of subdomains, which is an improvement from the standard GWRM $N_s^{1.4}$ scaling. 

The ideal MHD equations were solved within $0.1$\% error for the instability growth in a screw-pinch, which agrees well with previous simulations \cite{Scheffel:GWRM1}. It was able to reach  this accuracy using one time interval, making the CBC-GWRM highly competitive when solving for slow unstable mode growth rates. This shows that the CBC-GWRM is capable of solving complex physical systems that are relevant for fusion plasma physics, whilst efficiently resolving the temporal domain.

It is found that the CBC-GWRM is more efficient than the global GWRM if the ${\partial}x_p/{\partial}x_{BC}$ variables are computed numerically. This allows the CBC-GWRM to be more competitive when many subdomains are used. A clear example of this can be seen when comparing the GWRM subdomain schemes for the linearised Burger equation. Given the same parameters and sixty subdomains the CBC-scheme uses approximately 60\% less memory than the global subdomain scheme. More importantly, the CBC subdomain scheme can be parallelized so that the full potential of multiple CPUs and GPU acceleration can be harnessed. For large problems in fluid dynamics and MHD the CBC-scheme consists of 90-95\% of potentially parallelizable code, making high speedup gains possible.

\bibliographystyle{siamplain}
\bibliography{mybib}


%\begin{thebibliography}{mybib}

%\end{thebibliography}
\end{document}